\documentclass[journal]{IEEEtran}
\ifCLASSINFOpdf
\else
\fi
\usepackage{amsmath, amssymb}
\usepackage{color}
\usepackage{graphicx}
\usepackage{booktabs}
\usepackage{pifont}
\usepackage{listings}
\usepackage{graphicx}
\usepackage{caption}
\usepackage{subcaption}
\usepackage{xcolor}
\usepackage{multirow}
\usepackage{makecell}
\usepackage[linesnumbered,ruled,vlined]{algorithm2e}
\usepackage{enumitem}
\usepackage{tcolorbox}
\usepackage{threeparttable}
\usepackage{balance}
\usepackage{cite}
\usepackage{url}
\usepackage{hyperref}
\usepackage{wasysym}
\usepackage{mathpartir}
\usepackage{colortbl}
\usepackage{orcidlink}

\newcommand{\etal}{et al.}
\newcommand{\eg}{e.g.}

\newcommand{\ccSysName}[0]{DITING\xspace}
\newcommand{\ccBenchName}[0]{PartitioningE-Bench\xspace}
\newcommand{\ccTitle}[0]{\ccSysName: A Static Analyzer for Identifying Bad Partitioning Issues in TEE Applications}

\newcommand{\whiteding}[1]{\ding{\numexpr171+#1\relax}}

\begin{document}
\title{\ccTitle}

\author{Chengyan~Ma~\textsuperscript{\orcidlink{0000-0001-9256-6930}},
        Ruidong~Han,
        Jieke~Shi,
        Ye~Liu,
        Yuqing~Niu,
        Di~Lu~\textsuperscript{\orcidlink{0000-0002-9923-6405}},
        Chuang~Tian~\textsuperscript{\orcidlink{0000-0003-1723-7859}},
        Jianfeng~Ma~\textsuperscript{\orcidlink{0000-0003-4251-1143}},
        Debin~Gao~\textsuperscript{\orcidlink{0000-0001-9412-9961}},
        and~David~Lo~\textsuperscript{\orcidlink{0000-0002-4367-7201}},~\IEEEmembership{Fellow,~IEEE}% <-this % stops a space
\thanks{Chengyan~Ma, Ruidong~Han, Jieke~Shi, Ye~Liu, Yuqing~Niu, Debin~Gao and David~Lo are with the Centre on Security, Mobile Applications and Cryptography, Singapore Management University, Singapore 188065, Singapore (e-mail: chengyanma@smu.edu.sg, rdhan@smu.edu.sg, jiekeshi@smu.edu.sg, yeliu@smu.edu.sg, yuqingniu@smu.edu.sg, dbgao@smu.edu.sg, davidlo@smu.edu.sg).}% <-this % stops a space
\thanks{Di~Lu, Chuang~Tian and Jianfeng~Ma are with Xidian University, Xi'an 710071, China (e-mail: dlu@xidian.edu.cn, tianchuang@xidian.edu.cn, jfma@mail.xidian.edu.cn).}% <-this % stops a space
% \thanks{Manuscript received April 19, 2005; revised August 26, 2015.}
}


\markboth{Journal of \LaTeX\ Class Files,~Vol.~14, No.~8, August~2015}%
{Shell \MakeLowercase{\textit{et al.}}: Bare Demo of IEEEtran.cls for IEEE Journals}



\maketitle

\begin{abstract}
Trusted Execution Environment (TEE) enhances the security of mobile applications and cloud services by isolating sensitive code in the secure world from the non-secure normal world. However, TEE applications are still confronted with vulnerabilities stemming from bad partitioning. Bad partitioning can lead to critical security problems of TEE, such as leaking sensitive data to the normal world or being adversely affected by malicious inputs from the normal world.

To address this, we propose an approach to detect partitioning issues in TEE applications. First, we conducted a survey of TEE vulnerabilities caused by bad partitioning and found that the parameters exchanged between the secure and normal worlds often contain insecure usage with bad partitioning implementation. Second, we developed a tool named \ccSysName that can analyze data flows of these parameters and identify their violations of security rules we defined to find bad partitioning issues. Different from existing research that only focuses on malicious input to TEE, we assess the partitioning issues more comprehensively through input/output and shared memory. Finally, we created the first benchmark targeting bad partitioning, consisting of 110 test cases. Experiments demonstrate that \ccSysName achieves an F1 score of 0.90 in identifying bad partitioning issues.
\end{abstract}


% \fi
\IEEEpeerreviewmaketitle

\section{Introduction} \label{s:intro}
\IEEEPARstart{T}{rusted} Execution Environment (TEE) has emerged as an essential component in enhancing the security of mobile applications and cloud services~\cite{10738357, 10.1145/2541940.2541949, 9693116}.
In a typical TEE application, code is usually divided into secure and non-secure parts. The non-secure code operates in the non-secure environment, also known as the \textit{normal world}, while the secure code runs within a TEE.
This mechanism provides significant security advantages by isolating the secure code from the normal world, preventing attackers from accessing sensitive data or disrupting secure functions, such as key management~\cite{8359163, 9024053}, cryptography~\cite{10745433, mark2021rvtee}, and memory operations~\cite{DBLP:conf/ndss/ZhaoM19, 10.1145/3620665.3640378}.
To achieve this isolation, TEE leverages hardware-based features like memory partitioning and access control enforced by the CPU~\cite{10646815, 10.1002/cpe.4130}.
When the code in the normal world needs to access secure functions or data, it must interact with TEE using specific interfaces.
Therefore, partitioning strategies of TEE applications that incorporate accurate isolation~\cite{10.1145/3319535.3363205, 10.1145/2382196.2382214, DBLP:conf/ndss/KimKCGL0X18} and access control~\cite{9925586, 10.1145/2382196.2382214} mechanisms can be helpful in reducing the attack surface.
For instance, Zhao \etal~\cite{10.1145/3319535.3363205} designed SecTEE, a solution that integrates computing primitives (\eg, integrity measurement, remote attestation, and life cycle management) into TEE.
This design enables SecTEE to resist privileged host software attacks from the normal world through isolation.
Liu \etal~\cite{9925586} proposed a multi-owner access control scheme for access authorization in Intel SGX, which can prevent unauthorized data disclosure.
The above research has shown that well-designed partitions of TEE applications can help mitigate risks by establishing clear boundaries and secure communication protocols between the normal world and TEE.

However, TEE applications still face some threats. For example, Khandaker \etal~\cite{10.1145/3373376.3378486} summarized four attacks that can be implemented in Intel SGX: \textit{Concurrency}, \textit{Order}, \textit{Inputs}, and \textit{Nested} (COIN attacks).
In these attacks, malicious TEE users can exploit the lax detection vulnerabilities of non-secure inputs (\eg, buffer and value with incorrect size) in Intel SGX to interfere with executing programs in TEE.
Cerdeira \etal~\cite{9152801} conducted a security investigation on some popular systems based on TrustZone.
They found that input validation weaknesses and other bugs in TrustZone-assisted systems can be used to obtain permissions of TEE. For example, Huawei TEE allows a TEE application to dump its stack trace to the normal-world memory through debugging channels, leaking sensitive information.

\begin{figure}[t]
    \centering
    \includegraphics[width=0.98\linewidth]{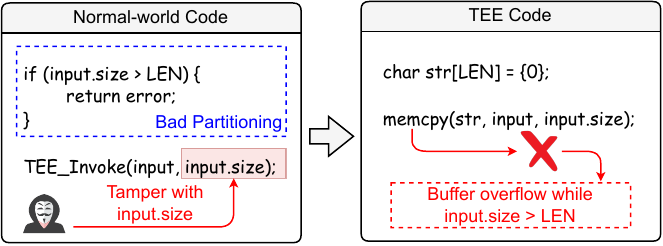}
    \caption{A bad partitioning case of input validation weaknesses.}
    \label{fig:bpc}
\end{figure}

Through analysis of these cases, we found that vulnerabilities in TEE applications often arise from insecure data communication between the normal world and TEE~\cite{10.1145/3407023.3407072, 8365755}, which usually relies on three types of parameters: input/output and shared memory~\cite{s20041090}.
These parameters are typically managed by the normal world, making them susceptible to unauthorized access.
Hence, the usage of these parameters without isolation and access control 
(\eg, cryptographic functions and bounds checking)
can introduce \textbf{bad partitioning} issues to TEE applications.
For example, there is a bad partitioning case in Fig.~\ref{fig:bpc}. The normal-world code validates the size of the input buffer before invoking TEE, but this validation step is missing inside the TEE code.
If an attacker controls the normal world and has the ability to intercept sensitive data or gain unauthorized access to secure operations~\cite{279916}, they can tamper with the size of the input buffer before the TEE invocation. The TEE application is then vulnerable to buffer overflow attacks caused by malicious inputs.
Such scenarios highlight the critical importance of ensuring that partition boundaries are well-defined and communication channels are fortified against potential breaches.
Therefore, this paper discusses TEE bad partitioning issues based on two key aspects: 
(1) the absence of secure interface components between the normal world and the TEE (\eg, returning output data without encryption, and using input data missing bounds checking), and (2) the use of insecure mechanisms within TEE code (\eg, insecure reliance on shared memory).
Unfortunately, there are currently no tools to analyze and identify TEE partitioning issues.

To address these challenges and enhance TEE security, we formulated three rules for detecting these issues. 
Following these rules, we designed a data flow analyzer to identify bad partitioning in TEE applications by determining whether input/output and shared memory parameters passed between the normal world and TEE violate the defined rules.
This is the first automatic tool that can identify bad partitioning issues in TEE applications.
Our contributions of this paper include:
\begin{itemize} [leftmargin=*]
    \item \textbf{The first systematic summary of partitioning issues.} 
    With the real-world vulnerable code where input, output, and shared memory parameters are handling in sensitive functions and data of TEE, we summarized that the bad partitioning issues in TEE are often caused by the improper usage of these parameters.
    Unlike previous research~\cite{10.1145/3373376.3378486, 9152801}, which primarily focused on malicious inputs to TEE, our approach addresses the potential vulnerabilities brought by three parameters within TEE, offering a more comprehensive solution to bad partitioning issues.
    \item \textbf{Proposing a parameter data flow analyzer to identify partitioning issues.} 
    Based on parameter types, we established a set of rules that these parameters must adhere to as they flow through sensitive functions and interact with sensitive data.
    By applying data flow analysis to determine whether these parameters violate the established rules caused by partitioning issues, we can effectively identify instances of bad partitioning in TEE applications.
    We named it \textbf{\ccSysName}, a fantastic beast that has the ability to distinguish between good and evil from the Chinese Mythology novel \textit{Journey to the West}.
    \item \textbf{Creating the benchmark for bad partitioning detection.} Due to the lack of datasets containing TEE partitioning issues, we created the first \textbf{Partitioning} \textbf{E}rrors \textbf{Bench}mark named \textbf{\ccBenchName}, which consists of 110 test cases, and 90 cases of them cover 3 types of bad partitioning vulnerabilities. Experimental results demonstrate that \ccSysName achieves an F1 score of 0.90 in identifying bad partitioning issues. Using \ccSysName, we also find 130 issues in 16 real-world TEE projects. Then, we have contributed some pull requests addressing the issues in 5 projects and one of them has already been confirmed and merged into the main branch by the maintainers.
\end{itemize}

The rest of the paper is organized as follows. We first describe the data interaction model and bad partitioning issues of TEE applications in Section~\ref{s:bg}. Next, Section~\ref{s:design} presents the system design of \ccSysName and Section~\ref{s:eva} evaluates the system. Subsequently, we discuss the threats to validity and related works in Section~\ref{s:val} and Section~\ref{s:related}, respectively. Finally, we conclude the paper and mention future work in Section~\ref{s:con}.

\section{Background and Motivation} \label{s:bg}
\subsection{TEE Data Interaction} \label{s:params}
\begin{figure}[t]
    \centering
    \includegraphics[width=0.75\linewidth]{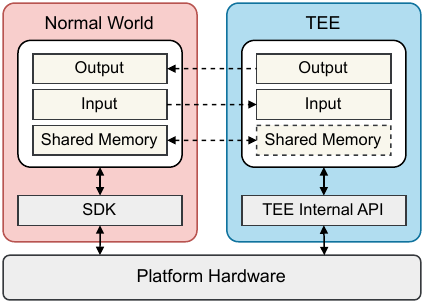}
    \caption{Data communication between TEE and the normal world.}
    \label{fig:datacom}
\end{figure}

As shown in Fig.~\ref{fig:datacom}, the normal world and TEE are two independent environments, separated to ensure the security of sensitive functions and data. In this architecture, the communication between TEE and the normal world involves three types of parameters: input, output, and shared memory~\cite{s20041090}.

Input parameters are used to transfer data from the normal side to TEE, while outputs handle the results or send processed data back. Inputs and outputs are simple mechanisms that allow users to temporarily transfer small amounts of data between the normal world and TEE. However, temporary inputs/outputs realize data transfer through memory copying, slightly lowering the performance of TEE applications due to additional memory copy. They are mainly used to transmit lightweight data, such as user commands and data for cryptographic operations.

Shared memory provides a zero-copy memory block to exchange larger data sets (\eg, multimedia files or bulk data) between two sides~\cite{optee}. It allows both sides to access the same memory space efficiently, which avoids frequent memory copying. Shared memory can also remain valid across different TEE invocation sessions, making it suitable for scenarios that require data to be reused multiple times.

Moreover, an SDK is responsible for managing these parameters and communication in the normal side. For example, TrustZone-based OP-TEE uses \texttt{TEEC\_InvokeCommand()} function to execute the in-TEE code, enabling the shared memory or temporary buffers to transfer data between the two sides~\cite{8684292}. 
Then, OP-TEE can handle these interactions through its internal APIs, which process incoming requests, perform secure computations, and return responses to the normal world.
Similarly, Intel SGX utilizes the ECALL and OCALL interfaces to achieve these functionalities~\cite{10632129}. ECALLs allow the normal world to securely invoke functions within the SGX enclave, passing data into the trusted environment for processing, while OCALLs enable the enclave to request services or share results with the untrusted normal world.

\subsection{Threat Model} \label{s:threat}
In this paper, we consider an adversary who has full control over the normal-world code and can tamper with the parameters passed between the normal world and TEE.
This includes the ability to manipulate input values, intercept or observe output data, and modify shared memory contents.
We identify three main threat vectors that emerge from bad partitioning practices:
\begin{itemize} [leftmargin=*]
    \item \textbf{Sensitive Data Leakage:} If sensitive data (\eg, cryptographic keys, passwords, or authentication tokens) is output to the normal world without encryption, it may be exposed to adversaries. 
    This risk is further exacerbated by inference attacks~\cite{10.1145/3707453}, where attackers infer protected information by observing output patterns or variations when they are returned in plain text.
    \item \textbf{Insecure Input Handling:} The TEE SDK typically performs basic validation checks when copying parameters into TEE, such as preventing null pointers. However, the SDK lacks semantic knowledge of how those inputs are used within the TEE application~\cite{10.1145/3373376.3378486}. Thus, the burden of input validation within TEEs, such as checking buffer sizes or ensuring value ranges, falls on the developer.
    If such validation is missing or incomplete, the TEE becomes vulnerable to memory security issues such as buffer overflows, even when inputs were superficially checked in the normal world.
    \item \textbf{Shared Memory Manipulation:} Both the normal side and TEE can access the shared memory.
    However, existing TEEs generally lack built-in mechanisms to ensure the integrity of shared memory contents~\cite{9309302}, and most hardware vendors do not provide standardized guidelines for secure shared memory usage~\cite{10.1145/3407023.3407072}.
    As a result, a common pitfall is that TEE applications may access shared memory directly through shallow copies, making them vulnerable to runtime manipulation by untrusted normal-world processes controlled by attackers. This risk has been documented in prior studies like EnclaveFuzz~\cite{DBLP:conf/ndss/ChenLMLC024} and prompted ongoing community discussions on best practices for securely handling shared memory in TEE systems\footnote{https://github.com/lsds/sgx-lkl/issues/745}.
\end{itemize}
The above vulnerabilities have also been exemplified in real-world TEE applications, such as CVE-2016-10237\footnote{https://www.cve.org/CVERecord?id=CVE-2016-10237}, CVE-2020-8936\footnote{https://www.cve.org/CVERecord?id=CVE-2020-8936} and CVE-2021-0186\footnote{https://www.cve.org/CVERecord?id=CVE-2021-0186}.
Therefore, our threat model focuses on attacks stemming from the misuse of input/output and shared memory parameters in the TEE code, while the misuse refers to the vulnerable code that misses secure components or directly calls insecure components when using the above parameters.

\begin{table*}[t]
    \caption{Surveys of bad partitioning issues in real-world TEE projects. \CIRCLE~means that we found the corresponding issue in this project, and \Circle~indicates that we did not find.}
    \label{tbl:project_list}
    \renewcommand{\arraystretch}{1.3}
    \footnotesize
    % \setlength{\tabcolsep}{2.5mm}
    % \resizebox{\linewidth}{!}{
    \centering
	\begin{tabular}{lp{4cm}<{\centering}p{4cm}<{\centering}p{4cm}<{\centering}}
		\toprule
            \textbf{Projects} & Unencrypted Data Output & Input Validation Weaknesses & Direct Usage of Shared Memory \\
		\midrule
            \rowcolor{lightgray}
            optee-sdp & \CIRCLE & \CIRCLE & \Circle \\
            optee-fiovb & \CIRCLE & \CIRCLE & \Circle\\
            \rowcolor{lightgray}
            basicAlg\_use & \CIRCLE & \CIRCLE & \Circle\\
            external\_rk\_tee\_user & \CIRCLE & \CIRCLE & \CIRCLE \\
            \rowcolor{lightgray}
            secvideo\_demo & \Circle & \Circle & \CIRCLE \\
            PPPL & \Circle & \CIRCLE & \CIRCLE \\
            \rowcolor{lightgray}
            acipher & \CIRCLE & \Circle & \Circle \\
            Lenet5\_in\_OPTEE & \CIRCLE & \CIRCLE & \Circle \\
            \rowcolor{lightgray}
            hotp & \CIRCLE & \Circle & \Circle \\
            random & \CIRCLE & \Circle & \Circle \\
		\bottomrule
	\end{tabular}
    % }
\end{table*}

\subsection{Bad Partitioning in TEE Projects} \label{s:bp}
We searched for TEE-related projects on GitHub using API keywords associated with TEE. 
After excluding TEE SDK projects, we selected the top 10 projects with the highest star counts.
By manually analyzing these TEE projects, we identified three primary types of bad partitioning issues: \whiteding{1} unencrypted data output, \whiteding{2} input validation weaknesses, and \whiteding{3} the direct usage of shared memory. 
We list some projects and issues they face in Table~\ref{tbl:project_list}, and discuss each of these issues in detail below.
In this paper, we use projects developed for ARM TrustZone as examples. The issues we identify and the proposed solutions work for other platforms as well.

\subsubsection{\textbf{Unencrypted Data Output}} \label{s:udo}
Since a TEE application can transfer data from TEE to the normal side using output parameters or shared memory, any unencrypted data is at risk of exposure to attackers during the output process, potentially compromising the confidentiality and security of sensitive data.

\begin{figure}[t]
  \centering
  \begin{subfigure}[b]{\linewidth}
    \begin{lstlisting}[language=c++]
char dump[MAX_DUMP_SIZE];
op.params[0].tmpref.buffer = (void *)dump;
op.params[0].tmpref.size = MAX_DUMP_SIZE - 1;
res = TEEC_InvokeCommand(&session, TA_SDP_DUMP_STATUS, &op, &err_origin);
    \end{lstlisting}
    \caption{The array \texttt{dump} is an output parameter in normal-world Code.}
  \end{subfigure}
  \hfill
  \begin{subfigure}[b]{\linewidth}
        \begin{lstlisting}[language=c++]
// params[0].memref is a memory copy of op.params[0].tmpref in the normal world
snprintf(params[0].memref.buffer, params[0].memref.size, "delta (decoder) refcount %d\n", delta_refcount);
// params[0].memref will be copied to op.params[0].tmpref after return
return TEE_SUCCESS;
    \end{lstlisting}
    \caption{Data \texttt{delta\_refcount} is directly copied to the output buffer in TEE code.}
  \end{subfigure}
  \caption{The example of unencrypted data output from optee-sdp.}
  \label{fig:bp1code}
\end{figure}

\begin{figure}[t]
  \centering
  \begin{subfigure}[b]{\linewidth}
    \begin{lstlisting}[language=c++]
char g_AesOutpUT[] = {0x01, 0x02, 0x03, 0x04, 0x05};
// op.params[1].tmpref is an input buffer smaller than test buffer in TEE
op.params[1].tmpref.buffer = g_AesOutpUT;
op.params[1].tmpref.size = 5;
res = l_CryptoVerifyCa_SendCommand(&op, &session, CMD_AES_OPER);
    \end{lstlisting}
    \caption{The input buffer and its size are defined in normal-world Code.}
  \end{subfigure}
  \hfill
  \begin{subfigure}[b]{\linewidth}
        \begin{lstlisting}[language=c++]
char test[256] = {};
// params[1].memref is the copy of op.params[1].tmpref in the normal world
TEE_MemMove(test, params[1].memref.buffer, params[1].memref.size);
    \end{lstlisting}
    \caption{\texttt{TEE\_MemMove} is called without verifying the size of input buffer in TEE code.}
  \end{subfigure}
  \caption{The example of input validation weaknesses from basicAlg\_use.}
  \label{fig:bp2code}
\end{figure}

\begin{figure}[t]
  \centering
  \begin{subfigure}[b]{\linewidth}
    \begin{lstlisting}[language=c++]
unsigned char membuf_input[] = "from_CA_to_TA";
// definition of shared memory
TEEC_SharedMemory shm = {
    .size = sizeof(membuf_input),
};
TEEC_AllocateSharedMemory(&ctx, &shm);
// shared memory can be changed outside TEE
memcpy(shm.buffer, membuf_input, shm.size);
op.params[2].memref.parent = &shm;
res = TEEC_InvokeCommand(&sess, TA_TEST_APP_FILL_MEM_BUF, &op, &err_origin);
    \end{lstlisting}
    \caption{Shared memory is managed in the normal world.}
  \end{subfigure}
  \hfill
  \begin{subfigure}[b]{\linewidth}
        \begin{lstlisting}[language=c++]
// params[2].memref is shared memory
void *buf = params[2].memref.buffer;
unsigned int sz = params[2].memref.size;
// params[2].memref may have been changed outside TEE
if (!TEE_MemCompare(buf, "from_CA_to_TA", sz)) {
    IMSG("membuf test : Pass!\n");
}
    \end{lstlisting}
    \caption{\texttt{TEE\_MemCompare} is invoked with shared memory in TEE code.}
  \end{subfigure}
  \caption{The example of the direct usage of shared memory from external\_rk\_tee\_user.}
  \label{fig:bp3code}
\end{figure}

Fig.~\ref{fig:bp1code} shows a scenario where if the TEE data is copied to an output buffer without being encrypted, then the attacker can directly obtain the plaintext from \texttt{dump} in the normal world. Therefore, implementing proper security components, such as cryptographic functions for encryption, is essential to prevent unauthorized access or data leakage during the TEE output.

\subsubsection{\textbf{Input Validation Weaknesses}} \label{s:ivw}
As mentioned in Section~\ref{s:threat}, an attacker with normal-world permissions has the ability to tamper with any data intended for input into TEE. Fig.~\ref{fig:bp2code} illustrates that even though the normal-world code defines a smaller buffer than \texttt{text} buffer used in TEE, it is still susceptible to buffer overflow attacks that in-TEE code directly performs memory copy operations in the TEE code without validating the size of the input buffer \texttt{params[1].memref.buffer}. This occurs because the TEE code may incorrectly assume that the input buffer is secure and reliable without explicitly checking its size. If an attacker maliciously alters the input buffer size, memory copy operations within TEE could exceed the boundaries of the inputs, leading to severe consequences such as crashes and data corruption. Similar precautions should be taken when using input values as indices for array access in TEE. 
We also call the above precautions security components, such as bounds checking for input parameters.

\subsubsection{\textbf{Direct Usage of Shared Memory}} \label{s:dusm}
Shared memory, as a zero-copy memory block, provides the same physical memory space accessible to both the secure and non-secure partitions. Since this memory is maintained by the normal-world code, directly using shared memory can introduce significant vulnerabilities. 
Fig.~\ref{fig:bp3code} shows that variable \texttt{buf} is initialized by a shallow copy of the shared memory. Therefore, the contents of \texttt{buf} can be directly modified from outside TEE. When \texttt{buf} is passed into \texttt{TEE\_MemCompare} function, the TEE application may not return the expected results. 
Therefore, we consider the code that directly uses shared memory to be an insecure component.

\begin{figure*}[t]
    \centering
    \includegraphics[width=0.9\textwidth]{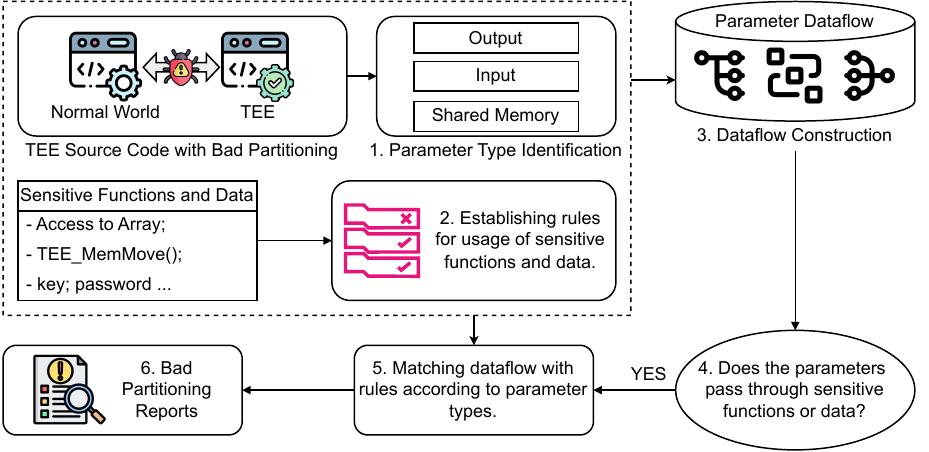}
    \caption{Overview workflow of \ccSysName.}
    \label{fig:workflow}
\end{figure*}

\section{System Design} \label{s:design}
In this section, we present the design of the proposed \ccSysName\footnote{https://github.com/CharlieMCY/PartitioningE-in-TEE} to identify bad partitioning issues of TEE projects. With the help of CodeQL~\cite{10.1002/spe.3199, codeql}, a code analysis engine developed by GitHub, we construct data flows of parameters passed between the normal world and TEE. Then, by analyzing the data flows, \ccSysName can detect insecure operations caused by bad partitioning. The design process involves 6 key steps shown in Fig.~\ref{fig:workflow}:

\textbf{Preparation Step: Rule Establishment.} Sensitive functions and data (\eg, array access, memory operations and key mentioned in Fig.~\ref{fig:workflow}) are critical to security and privacy. For example, sensitive data should not be directly passed to the normal world through output parameters. Rules are established to define permissible operations on these sensitive functions and data.

\textbf{Step 1: Parameter Type Identification.} The initial step is identifying the parameters exchanged between the normal world and TEE. These parameters are classified into three types based on Section~\ref{s:params}.

\textbf{Step 2: Data flow Generation.} Data flows of input/output and shared memory parameters will be generated. This will trace the path of each parameter and map its interactions with different components.

\textbf{Step 3: Parameter Analysis.} The next step involves analyzing whether the parameters pass through functions or interact with data related to rules in the preparation step.

\textbf{Step 4: Rule Matching with Parameter Data flows.} If a parameter passes through rule-related components, we match its data flow against the established rules in the preparation step to identify violations. Parameters that violate the rules will pinpoint bad partitioning issues, such as leakage of sensitive data to the normal world and misuse of shared memory leading to unexpected execution of in-TEE code.

\textbf{Step 5: Reporting Bad Partitioning Issues.} Based on the rule violations detected in the previous step, \ccSysName generates detailed reports highlighting bad partitioning issues.

By following this process, \ccSysName ensures robust detection of partitioning flaws in TEE projects, aiding developers in adhering to security policies and mitigating potential vulnerabilities. In the following, we will describe the implementation of \ccSysName in detail.

\subsection{Rule Establishment} \label{s:rules}

\begin{table}[t]
    \caption{Defined notations of parameters, variables, and operations in the TEE code.}
    \label{tbl:symbol}
    \renewcommand{\arraystretch}{1.3}
    \footnotesize
    \setlength{\tabcolsep}{1mm}
    \centering
	\begin{tabular}{|lcp{5.8cm}|}
            \hline
            $o$ & : & The output parameter \\
            $i$ & : & The input parameter\\
            $m$ & : & The shared memory parameter\\
            $x,y,z$ & : & The data in the TEE code\\
            $y:=\texttt{ENC}(x)$ & : & $y$ is a return value or buffer of the encryption-related cryptographic API, where $x$ is passed into. It can also be said that $x$ is encrypted into cipher text $y$\\
            $\texttt{COPY}(d, s)$ & : & Copying memory from the source buffer $s$ to destination buffer $d$\\
            % $o:=x$ & : & An assignment statement where $o$ obtains $x$\\
            $\texttt{IF}(i)$ & : & $i$ exists in the expression of a conditional statement\\
            $\texttt{ARRAY}(i)$ & : & An $i$-related array access\\
            $x:=\texttt{SHALLOW}(m)$ & : & $x$ is the shallow copy of the shared memory parameter $m$\\
            $x:=\texttt{MALLOC}(i)$ & : & An in-TEE buffer $x$ is allocated according to the input parameters $i$ \\
            $y:=\texttt{OP}(x)$ & : & Other operations besides the above, including arithmetic, function passing... \\
            \hline
	\end{tabular}
\end{table}

Given the discussion on bad partitioning in Section~\ref{s:bp}, we need to define a set of rules to detect these issues. These rules will focus on identifying insecure parameter interactions with in-TEE sensitive functions and data, which commonly arise from bad partitioning.
As shown in Table~\ref{tbl:symbol}, we use lowercase letters ($o$, $i$, $m$, $x$...) to denote parameters and data in the TEE code.
Meanwhile, we define some commonly used operations in the TEE code, such as data encryption functions $y:=\texttt{ENC}(x)$, conditional statements $\texttt{IF}(i)$, access to array $\texttt{ARRAY}(i)$, shallow copying $x:=\texttt{SHALLOW}(m)$, etc.
In addition, it should be noted that $\texttt{COPY}(d, s)$ is a deep copy function, which has two situations: output to the outside of TEE $\texttt{COPY}(o, x)$ and copying data into TEE $\texttt{COPY}(x, i)$.

Based on these notations, we provide the rules in Fig.~\ref{fig:full-inference-rules} and use the predicates in Table~\ref{tbl:predicate} for generating the data flows and detecting the issues.
Firstly, we give the propagation rules to represent the transfer path of in-TEE data or parameters.
$x \rightsquigarrow y$ denotes direct propagation, indicating that $y$ is directly derived from $x$, for example, $x$ is passed to $y$ through a function call or assignment statement. Then, $x \rightsquigarrow^* z$ denotes a transitive propagation, meaning that $x$ has undergone multiple propagations to $z$.

\begin{figure*}[t]
\centering
\small

% \begin{minipage}[t]{\linewidth}
\centering
\textbf{Propagation Rules}
\begin{mathpar}
\inferrule[Direct Propagation]
{ y := \texttt{OP}(x) }
{ x \rightsquigarrow y }

\inferrule[Transitive Propagation]
{ x \rightsquigarrow y \\ y \rightsquigarrow z }
{ x \rightsquigarrow^* z }
\end{mathpar}

\vspace{0.5em}
\textbf{Rule 1 for Unencrypted Data Output}
\begin{mathpar}
\inferrule[Source]
{ x }
{ \texttt{Source}(x) }

\inferrule[EncStmt]
{ y := \texttt{ENC}(x) }
{ \texttt{SC}(x) }

\inferrule[Sink]
{ y \rightsquigarrow^* z \\ \texttt{COPY}(o,z)}
{ \texttt{Sink}(o) }

\inferrule[Violation]
{ \texttt{Sink}(o) \\ \nexists \texttt{SC}(x) }
{ \texttt{BP} }
\end{mathpar}
% \end{minipage}

\vspace{0.5em}
% \begin{minipage}[t]{\linewidth}
\centering
\textbf{Rule 2 for Input Validation Weaknesses}
\begin{mathpar}

\inferrule[Source]
{ i }
{ \texttt{Source}(i) }

\inferrule[IfStmt]
{ \texttt{IF}(i) }
{ \texttt{SC}(i) }

\inferrule[Sink]
{ i \rightsquigarrow^* i' \\ \texttt{CS}\in\{\texttt{ARRAY}(i'),\texttt{COPY}(x,i')\} }
{ \texttt{Sink}(\texttt{CS})}

\inferrule[Violation]
{ \texttt{Sink}(\texttt{CS}) \\ \nexists \texttt{SC}(i) }
{ \texttt{BP} }
\end{mathpar}

\vspace{0.5em}
\textbf{Rule 3 for Direct Usage of Shared Memory}
\begin{mathpar}
\inferrule[Source]
{ m }
{ \texttt{Source}(m) }

\inferrule[ShallowCopyStmt]
{ x := \texttt{SHALLOW}(m) }
{ \texttt{ISC}(x) \\ \texttt{ISC}(m) }

\inferrule[Sink]
{  m \rightsquigarrow^* y \\  x \rightsquigarrow^* y }
{ \texttt{Sink}(y) }

\inferrule[Violation]
{ \texttt{Sink}(y) \\ \exists\texttt{ISC}(m) \lor \exists\texttt{ISC}(x) }
{ \texttt{BP} }
\end{mathpar}
\caption{Rules for detecting bad partitioning (\texttt{BP}) issues by checking the violation usage of TEE parameters.
The violations include two types: (1) sensitive statements or parameter usage that lack security components, (2) direct usage of insecure components.}
\label{fig:full-inference-rules}
\end{figure*}

\begin{table}[t]
    \caption{Predicates used in the static analysis. $*$ denotes a node in the data flow, and the predicate presents the state of a node based on the rules.}
    \label{tbl:predicate}
    \renewcommand{\arraystretch}{1.3}
    \footnotesize
    \centering
	\begin{tabular}{cl}
            \toprule
            \textbf{Predicates} & \textbf{Description} \\
            \midrule
            $\texttt{Source}(*)$ & $*$ is a source node of the data flow \\
            $\texttt{Sink}(*)$ & $*$ is a sink node of the data flow \\
            $\texttt{SC}(*)$ & $*$ is a secure component \\
            $\texttt{ISC}(*)$ & $*$ is a insecure component \\
            $\texttt{CS}$ & A critical statement \\
            $\texttt{BP}$ & A bad partitioning issue\\
            \bottomrule
	\end{tabular}
\end{table}

In Rule 1, we assume that $x$ is an in-TEE data that is eventually passed to an output parameter. If $x$ is not passed into an encryption API (\eg, \texttt{TEE\_CipherUpdate()}), it is treated as unencrypted (i.e., plain text).
Consequently, any output parameters that receive $x$ via memory copy operations (e.g., $\texttt{COPY}(o,x)$) are identified as bad partitioning instances.

In Rule 2, we address the lack of bounds checking of the input parameters that are used in array accesses or memory copy operations.
Specifically, an input value may be an index of an array (\eg, \texttt{a[params[0].value.a]}), or an input buffer can also be accessed by any index (\eg, \texttt{params[1].memref.buf[n]}).
In such cases, it is necessary to ensure that the index does not exceed the valid bounds of the array or buffer.
Similarly, in the memory copy operation $\texttt{COPY}(x,i')$, the size of the input buffer should be smaller than the size of the destination buffer $x$; otherwise, a buffer overflow may occur.
Therefore, array accesses and memory copies that are not preceded by conditional checks (i.e., $\texttt{IF}(i)$) are flagged as bad partitioning instances.
An exception to this rule occurs when the TEE application explicitly allocates memory whose size matches that of the input buffer. In such cases, a direct copy operation without a conditional check is considered safe, as shown below:
\begin{mathpar}
\inferrule[]
{ i\rightsquigarrow^* i' \\ \texttt{COPY}(x,i') \\ \exists x:=\texttt{MALLOC}(i') }
{ \neg \texttt{BP} }
\end{mathpar}

Rule 3 captures a scenario where shallow copying shared memory allows both the TEE and the normal world to access the same physical memory region. This violates the isolation guarantee of TEE and is therefore identified as a bad partitioning instance under our threat model.

To facilitate the detection of bad partitioning issues, we further refine the preceding rules by introducing a fine-grained classification of TEE operations. Specifically, memory copy and array access operations are defined as critical statements (\texttt{CS}), while encryption functions and conditional checks are classified as secure components (\texttt{SC}). 
Meanwhile, the TEE code that directly accesses normal-world memory (i.e., shared memory) is an insecure component (\texttt{ISC}).
Based on this classification: 
\begin{itemize}
\item Violations of Rules 1 and 2 can be detected by identifying critical statements that lack an associated secure component;
\item Violations of Rule 3, which involve the direct use of shared memory or its shallow copies, can be identified by detecting the presence of an insecure component.
\end{itemize}

\subsection{Data flow Generation} \label{s:dfg}
In this step, we construct the data flow of each parameter passed between the normal world and TEE according to the source and sink shown in Fig.~\ref{fig:full-inference-rules}.
For the data flow of output parameters, the source is the in-TEE data, and the sink is an output parameter, which is the destination buffer for a memory copy function or the left value of an assignment statement.
For input parameters, the source is the input value provided by the normal world, and the sink includes input-related operations such as array accesses and memory copies.
For shared memory, the data flow encompasses all operations that directly access shared memory or its shallow copies within the TEE code.

\subsection{Detection of Bad Partitioning}
\begin{algorithm}[t]
\caption{Detection of bad partitioning issues based on the data flow.}
\label{algo:df}
\footnotesize
\SetKwFunction{isSecure}{isSecureComponent}
\SetKwFunction{isInsecure}{isInsecureComponent}
\SetKwFunction{isCriticalStmt}{isCriticalStmt}
\KwIn{$DF$, the data flow of TEE parameter}
\KwOut{$R$, the report of bad partitioning issues}
$isBP \gets True$\;
\ForEach{$node \in DF$}{
    \If{\isSecure{$node$}}{
        $isBP \gets False$\;
    }
    \If{\isInsecure{$node$}}{
        $R \gets R \cup \{node\}$\;
    }
    \If{\isCriticalStmt{$node$} and $isBP$}{
        $R \gets R \cup \{node\}$\;
    }
}
\end{algorithm}
Based on the rules in Section~\ref{s:rules} and data flow in Section~\ref{s:dfg}, we can detect bad partition issues by traversing the data flow graph and checking rule violations at each relevant node.
As shown in Algorithm~\ref{algo:df}, each $node$ in the data flow $DF$ presents an operation statement, and all nodes are ordered according to program execution.
A node is flagged as a bad partitioning instance if the node is a critical statement that is not preceded by a corresponding secure component as required by the rules. 
Additionally, any insecure component (e.g., direct use of shared memory) is immediately reported as a violation. 
Since the data flow may involve different types of parameters (input, output, or shared memory), each detected issue is recorded in the relevant report $R$ according to its associated parameter type.

\section{Evaluations} \label{s:eva}
We have conducted several experiments to prove the effectiveness of \ccSysName. In particular, our evaluation focuses on addressing the following research questions (RQs):

\begin{enumerate}[label=\textbf{RQ\arabic*.}, leftmargin=1cm]
    \item How effective is \ccSysName?
    \item How efficient is \ccSysName?
    \item How \ccSysName performs for the real-world TEE projects?
\end{enumerate}

\subsection{Environment Setup}
\subsubsection{Benchmark for Evaluation} \label{s:bench}
Prior studies~\cite{10.1145/3540250.3549128, 8901573} have created benchmarks for various static analysis tools that involve injecting vulnerabilities into code.
However, none of the existing studies have created a benchmark for TEE partitioning issues.
By referencing the construction methodology of the popular API-misuse dataset CryptoAPI-Bench~\cite{8901573, 9721567, 10225251}, we designed the benchmark \ccBenchName for evaluating the abilities of \ccSysName in bad partitioning detection.
We will describe its creation steps below.

\begin{figure}[t]
  \centering
  \begin{subfigure}[b]{\linewidth}
    \begin{lstlisting}[language=c++]
int function(TEE_Param params[4])
{...
    params[0].value.a = var;
    ...
}
    \end{lstlisting}
    \caption{Example code of the in-procedure cases. In this case, the critical data \texttt{params} may be passed to another in-TEE function instead of being used directly after entering TEE, making it more challenging to track its data flow.}
    \label{f:eic}
  \end{subfigure}
  \hfill
  \begin{subfigure}[b]{\linewidth}
        \begin{lstlisting}[language=c++]
snprintf(params[1].memref.buffer, params[1].memref.size, "%s-%s-%d", key, vi, var);
    \end{lstlisting}
    \caption{Example code of the variadic function case. In this case, \texttt{snprintf} may have multiple arguments that need to be copied to the output buffer. Each argument must be checked to ensure it is encrypted.}
    \label{f:efs}
  \end{subfigure}
  \begin{subfigure}[b]{\linewidth}
        \begin{lstlisting}[language=c++]
if(params[2].memref.size > size) {
    memcpy(params[2].memref.buffer, buf, size);
}
    \end{lstlisting}
    \caption{Example code of control-flow-based cases. In this case, the memory operation \texttt{memcpy} is inside a conditional statement that checks the buffer size, so this operation is secure.}
    \label{f:eps}
  \end{subfigure}
  \caption{Examples of test cases in \ccBenchName.}
  \label{fig:cases}
\end{figure}

We first collected code snippets with known bad partitioning issues from real-world TEE projects mentioned in Section~\ref{s:bp} to construct the basic dataset and further expanded the dataset following the methodology in~\cite{8901573}.
We then categorized the test cases into 5 groups, including basic cases, in-procedure cases, variadic function cases, control-flow-based cases, and combined cases. 
These categories cover a wide range of bad partitioning patterns and enable comprehensive evaluation of our detection approach.
\begin{enumerate} [leftmargin=*]
    \item \textbf{Basic cases} are some simple instances described in in Section~\ref{s:bp}. These vulnerabilities typically occur in statements where parameters are directly used after invoking TEE.
    \item \textbf{In-procedure cases} mean that the parameter may be passed into other procedures or methods as an argument, which is shown in Fig.~\ref{f:eic}.
    \item Fig.~\ref{f:efs} gives a \textbf{variadic function case}, in which \texttt{snprintf()} may have more than one argument copied to the output buffer. We need to perform data flow analysis on each of them to find the issues of unencrypted data output.
    \item \textbf{Control-flow-based cases} will provide some \textbf{no-issue cases}, which can evaluate the precision of \ccSysName. For example, in Fig.~\ref{f:eps}, line 2 should not be identified as a bad partitioning issues, because it is in an \texttt{if} block, which checks the buffer size.
    \item \textbf{Combined cases} will introduce at least two of the above cases.
\end{enumerate}

As illustrated in Table~\ref{tbl:gttn}, we totally provide 110 cases for \ccBenchName, which are written in C programming language and developed for ARM TrustZone. Among these cases, 90 cases contain bad partitioning issues and 20 cases have no issues.

\begin{table}[t]
    \caption{The number of bad partitioning issues and no-issue cases in \ccBenchName.}
    \label{tbl:gttn}
    \renewcommand{\arraystretch}{1.3}
    \footnotesize
    \centering
	\begin{tabular}{lc}
		\toprule
		\makecell[c]{\textbf{Bad Partitioning Issues}} & \textbf{Number of Cases}\\ 
            \midrule
            Unencrypted Data Output & 35 \\
            Input Validation Weaknesses & 29 \\
            Direct Usage of Shared Memory & 26 \\
            \midrule
            \makecell[c]{\textbf{No-issue Cases}} & 20 \\
            \midrule
            \makecell[c]{\textbf{Total Cases}} & 110 \\
		\bottomrule
	\end{tabular}
\end{table}

\subsubsection{Experimental Platform}
\ccSysName is implemented by Python in combination with the static analysis tool CodeQL. CodeQL is utilized to generate the data flow for each parameter based on sources and sinks in Section~\ref{s:dfg}. The Python code is responsible for retrieving these data flows and identifying the locations of bad partitioning issues with the help of Algorithms~\ref{algo:df}.
All experiments were conducted on a machine running Ubuntu 24.04, equipped with a 48-core 2.3GHz AMD EPYC 7643 processor, and 512GB RAM.

\subsection{RQ1: Precision and Recall Evaluation}

\begin{table}[t]
    \caption{The detection precision and recall of \ccSysName on \ccBenchName. NI is the number of partitioning issues, N is the number of detection results, and TP is the number of true positives. P, R and F1 indicates the precision, recall, and F1 score, respectively\textsuperscript{*}.
    }
    \label{tbl:recall_res}
    \renewcommand{\arraystretch}{1.3}
    \footnotesize
    \setlength{\tabcolsep}{1.5mm}
    \centering
    \begin{threeparttable}
        \begin{tabular}{lcccccc}
        \toprule
        \makecell[c]{\textbf{Bad Partitioning Issues}} & \textbf{NI} & \textbf{N} & \textbf{TP} & \textbf{P(\%)} & \textbf{R(\%)} & \textbf{F1}\\
        \midrule
        Unencrypted Data Output & 35 & 34 & 33 & 97.06 & 94.29 & 0.96 \\
        Input Validation Weaknesses & 29 & 27 & 23 & 85.19 & 79.31 & 0.82 \\
        Direct Usage of Shared Memory & 26 & 30 & 25 & 83.33 & 96.15 & 0.89 \\
        \midrule
        \makecell[c]{\textbf{Total}} & 90 & 91 & 81 & 89.01 & 90 & 0.90\\
        \bottomrule
        \end{tabular}
    \begin{tablenotes}
        % \footnotesize
        \item[*] Note: $P(\%) = TP / N \times 100$, $R(\%) = TP / NI \times 100$.
    \end{tablenotes}
    \end{threeparttable}
\end{table}

Table~\ref{tbl:recall_res} lists the ground truth count of bad partitioning instances on \ccBenchName, alongside the number of issues reported by \ccSysName. It can be seen that \ccSysName achieves the highest precision of 97.06\% in detecting bad partitioning of unencrypted data output. However, for the other two issues, more FPs are observed.

Through analysing the examples of misdiagnosed code shown in Fig.~\ref{fig:fp}, we categorize the FPs into the following three types:

\begin{figure}[t]
  \centering
    \begin{subfigure}[b]{\linewidth}
    \begin{lstlisting}[language=c++]
char buf[] = "aabbcc";
...
params[0].memref.size = strlen(buf);
// params[1].value.a is the output, while params[2].value.a is the input
params[1].value.a = params[2].value.a;
    \end{lstlisting}
    \caption{The length of \texttt{buf} and \texttt{params[2].value.a} are not the sensitive data.}
    \label{fig:fp1}
  \end{subfigure}
  \begin{subfigure}[b]{\linewidth}
    \begin{lstlisting}[language=c++]
unsigned int size = params[0].memref.size;
// obtain the third last character in the input buffer
char c = params[0].memref.buffer[size - 3];
    \end{lstlisting}
    \caption{Line 3 is the FP of input validation weaknesses, where \texttt{size} is an input parameter as array index.}
    \label{fig:fp2}
  \end{subfigure}
  \hfill
  \begin{subfigure}[b]{\linewidth}
    \begin{lstlisting}[language=c++]
// variable assignment by shared memory value
unsigned int size = params[3].memref.size;
// shallow copy of shared memory buffer
void *buf = params[3].memref.buffer;
    \end{lstlisting}
    \caption{Line 2 is the FP of direct usage of shared memory, where \texttt{params[0].memref.size} is a shared memory parameter.}
    \label{fig:fp3}
  \end{subfigure}
  \caption{Examples of FPs in \ccSysName reports.}
  \label{fig:fp}
\end{figure}

\begin{figure}[t]
  \centering
  \begin{subfigure}[b]{\linewidth}
    \begin{lstlisting}[language=c++]
void test(char *dest, char *src)
{
    TEE_MemMove(dest, src, strlen(src));
}

void TA_InvokeCommandEntryPoint()
{
    char key[] = "123456";
    ...
    char *str = TEE_Malloc(strlen(key) + 1, 0);
    // case 1
    test(str, key);
    ...
    // case 2
    test(params[1].memref.buffer, key);
}
    \end{lstlisting}
    \caption{Line 15 is the FN of unencrypted data output, where \texttt{params[1].memref.buffer} is a raw pointer passed into \texttt{test}.}
    \label{fig:fn1}
  \end{subfigure}
  \hfill
  \begin{subfigure}[b]{\linewidth}
    \begin{lstlisting}[language=c++]
char *str[1024] = {0};
...
for (int i = 0; i < params[2].memref.size; i++) {
    str[i] = params[2].memref.buffer[i];
}
    \end{lstlisting}
    \caption{Line 4 is the FN of input validation weaknesses, where \texttt{i} is affected by the input \texttt{params[2].memref.size}.}
    \label{fig:fn2}
  \end{subfigure}
  \caption{Examples of FNs in \ccSysName reports.}
  \label{fig:fn}
\end{figure}

\begin{enumerate} [leftmargin=*]
    \item \textbf{Non-sensitive data is assigned to the output:} Some data can be assigned to the output without encryption, and this will not compromise the confidentiality of TEE. For example, in Fig.~\ref{fig:fp1}, it is acceptable that we pass the length of the output buffer to the outside or copy the input directly to the output.
    \item \textbf{Indexing for Input Buffer with Its Size:} As shown in Fig.~\ref{fig:fp2}, although the index of array access is influenced by the input value \texttt{params[0].memref.size}, TEE can ensure that this input will not lead to buffer overflow. This is because the input \texttt{params[0].memref.buffer} in TEE is actually a memory copy of the normal-side data, which is allocated with a size based on the above input value, and this input value is usually the size of the input data outside TEE. It means that the size of \texttt{params[0].memref.buffer} must not be smaller than the input value \texttt{params[0].memref.size}. Therefore, it is secure to directly use \texttt{params[0].memref.size} as an array index, even if no checks are performed on it which violates Rule 2.
    \item \textbf{Variable Assignment Using Shared Memory:} Similar to the input buffer, shared memory also transmits memory address and memory size, from the normal side to TEE by \texttt{params[3].memref}, thus, both of these parameters are treated as the type of shared memory. Meanwhile, any assignment statement containing shared memory parameters is not allowed in Rule 3. However, the variable assignment in line 2 of Fig.~\ref{fig:fp3} can be acceptable, as it differs from the insecure shallow copy of shared memory in line 4.
\end{enumerate}

According to the Recall rate in Table~\ref{tbl:recall_res}, \ccSysName can give more comprehensive coverage of bad partitioning related to the direct use of shared memory than other issues.
% In order to analyze the missed detection, we get FNs by excluding true positives (TP) from GTs.
We also give two examples of FNs in Fig.~\ref{fig:fn} and discuss them as follow:

\begin{enumerate} [leftmargin=*]
    \item \textbf{In-procedure Cases with Raw Pointers:} 
    Different from the parameter passing containing \texttt{TEE\_Param} type in Fig.~\ref{f:eic},
    % \texttt{params[1].memref.buffer} in line 15 of Fig.~\ref{fig:fn1} is a raw pointer passed to \texttt{dest}, which is a formal parameter of function \texttt{test}.
    line 15 of Fig.~\ref{fig:fn1} passes a raw pointer (\texttt{params[1].memref.buffer}) to \texttt{dest}, a formal parameter of the function \texttt{test}.
    Due to the loss of type semantics during raw pointer passing, \ccSysName detects the copy of sensitive data \texttt{key} to \texttt{dest} but does not flag it as a violation, as it cannot identify \texttt{dest} as an output parameter without its type.
    Additionally, in line 12, the copy of \texttt{key} to a local buffer \texttt{str} (defined within the TEE code) via the same function is considered secure.
    These in-procedure cases involving raw pointers are therefore difficult to detect due to the lack of type semantics.
    \item \textbf{Array Access Affected by Input:} Lines 3 to 5 give a \texttt{for} loop block, which is responsible for copying an buffer byte by byte.
    At this point, the input value \texttt{params[2].memref.size} does not directly serve as an index for array access but is instead used in the condition of \texttt{for} statement to constrain the value of \texttt{i}.
    Therefore, \ccSysName does not recognize this type of array access statement, that has broken Rule 2.
\end{enumerate}

In summary, if we formulate additional filters for the reported issues in the future, it is possible to effectively reduce FPs that deviate from the predefined rules but do not actually introduce security risks. What's more, we can refine the rules by incorporating the semantic context of the code, and FNs can be further reduced.

\begin{tcolorbox} [colback=gray!20!white]
\textbf{Answer to RQ1:}
(1) \ccSysName provides the highest F1 score of 0.96 in the detection of unencrypted data output (both precision and recall exceed 90\%).
(2) The detection of direct usage of shared memory achieves the best recall (above 95\%), but with relatively low precision (below 85\%).
(3) The detection of input validation weaknesses has an F1 score of 0.82. While its precision is above 85\%, the recall remains slightly lower at approximately 79\%.
\end{tcolorbox}

\subsection{RQ2: Efficiency Evaluation}
On our benchmark tests, we recorded the time taken by \ccSysName to generate data flows for three types of parameters and match these data flows against the corresponding rules.

\begin{figure}
    \centering
    \includegraphics[width=0.8\linewidth]{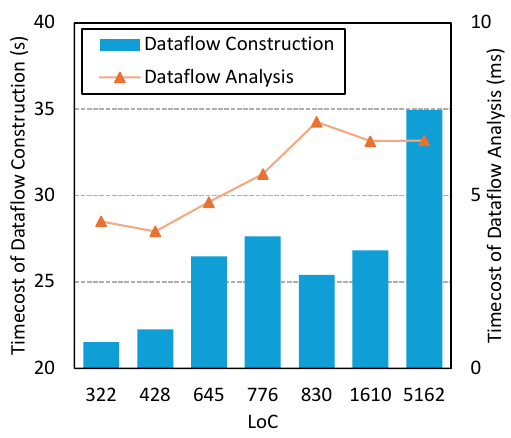}
    \caption{Time cost of data flow construction and analysis under different scales of LoC in \ccSysName.}
    \label{fig:effi}
\end{figure}

As shown in Fig.~\ref{fig:effi}, the time consumed by constructing data flows increases from 21.53 seconds to 34.95 seconds as the lines of code (LoC) grow. 
Then, \ccSysName focuses only on critical functions and data during data flow analysis, such as \texttt{TEE\_MemMove} and array access operations. The time spent on data flow analysis remains relatively low and does not grow significantly with the changes of LoC (from 3.96 milliseconds to 7.13 milliseconds).

\begin{tcolorbox} [colback=gray!20!white]
\textbf{Answer to RQ2:}
As the code scale increases from 322 LoC to 5162 LoC, the time cost of the data flow construction for bad partitioning analysis increases from 21.53 seconds to 34.95 seconds.
However, this time consumption also depends on the performance of the underlying code analysis tool.
\ccSysName still exhibits high efficiency for detecting bad partitioning issues in TEE projects (around 5.56 milliseconds).
\end{tcolorbox}

\subsection{RQ3: In-the-Wild Study}
Beside 10 real-world TEE projects in Section~\ref{s:bp}, we also select another 6 projects to evaluate the performance of \ccSysName on real-world programs, as shown in Table~\ref{tbl:expr_res}. 
These projects include 10 projects that have been manually analyzed in Section~\ref{s:bp}, along with 6 additional projects.

In these projects, \ccSysName reports 154 issues caused by bad partitioning, and 130 of them are confirmed by ourselves.
For example, in the \textit{optee-sdp} project (an application for securely storing device data in the TEE registers), the log functionality implemented via \texttt{snprintf} copies a large amount of register data into the output parameters, which are accessible only to TEE. This compromises the confidentiality of TEE.
Additionally, in projects such as \textit{Lenet5\_in\_OPTEE} (a Lenet5 convolutional neural network application in ARM TrustZone) and \textit{external\_rk\_tee\_user} (a sanity testsuite for OP-TEE), there are instances where input buffers are directly used in the memory copy functions without validating their length.
In particular, the \textit{PPFL} project (a privacy-preserving federated learning framework with TEE) uses a large number of constants as the access indexes of the input buffer, but lacks the validation of whether the input buffer size satisfies the constant value. Therefore, \textit{PPFL} has a large number of issues that violate Rule 2.
Attackers could exploit the issues of input validation weaknesses to cause buffer overflows in TEE applications.
Then, in the \textit{secvideo\_demo} project (a proof-of-concept for secure video playback on ARM TrustZone hardware), encryption and decryption operations are directly performed on video data in shared memory, undermining the isolation provided by TEE.
What's more, we have contributed five pull requests addressing those involving input validation weaknesses, which are the most feasible to fix without requiring extensive architectural changes. One of these patches\footnote{https://github.com/shuaifengyun/basicAlg\_use/pull/1} has already been confirmed and merged into the main branch by the maintainers.

\begin{table*}[t]
    \caption{The performance of \ccSysName in detecting three kinds of bad partitioning issues on real-world TEE projects. 
    }
    \label{tbl:expr_res}
    \renewcommand{\arraystretch}{1.3}
    \footnotesize
    \setlength{\tabcolsep}{2.5mm}
    % \resizebox{\linewidth}{!}{
    \centering
	\begin{tabular}{lr cc | cc | cc}
		\toprule
		\multirow{2}{*}{\textbf{Projects}} & 
            \multirow{2}{*}{\textbf{LoC}} & 
            \multicolumn{2}{c}{Unencrypted Data Output} & 
            \multicolumn{2}{c}{Input Validation Weaknesses} & \multicolumn{2}{c}{Direct Usage of Shared Memory} \\
            \cline{3-8}
             & & Report Number & True Positive & 
                 Report Number & True Positive & 
                 Report Number & True Positive \\
		\hline
            % synthesis & 515 & 8 & 8 & 0 & 100 & 
            %             6 & 5 & 1 & 83.33 & 
            %             6 & 5 & 1 & 83.33 \\
            \rowcolor{lightgray}
            optee-sdp & 830 & 11 & 11 &
                               7 &  7 & 
                             --- &  --- \\
            optee-fiovb & 776 & 2 & 1 & 
                                2 & 2 & 
                              --- & --- \\
            \rowcolor{lightgray}
            basicAlg\_use & 5,162 & 2 & 2 &  
                                    1 & 1 &  
                                  --- & --- \\
            external\_rk\_tee\_user & 716 & 4 & 2 & 
                                            2 & 2 & 
                                            1 & 1 \\
            \rowcolor{lightgray}
            secvideo\_demo & 645 & --- & --- & 
                                   --- & --- &
                                     7 & 4 \\
            % darknetz & 41,608 & --- & --- &
            %                     77 & --- &
            %                      1 & 1 \\
            PPFL & 80,589 & --- & --- &
                                86 & 75 &
                                 1 & 1 \\
            \rowcolor{lightgray}
            acipher & 338 & 1 & 0 &
                          --- & --- &
                          --- & --- \\
            Lenet5\_in\_OPTEE & 1,610 & 1 & 1 & 
                                        3 & 2 &
                                      --- & --- \\
            \rowcolor{lightgray}
            hotp & 428 & 1 & 1 & 
                       --- & --- &
                       --- & --- \\
            random & 322 & 1 & 1 & 
                       --- & --- &
                       --- & --- \\
            \rowcolor{lightgray}
            read\_key & 404 & 2 & 1 & 
                       --- & --- &
                       --- & --- \\
            save\_key & 529 & 2 & 1 & 
                       --- & --- &
                       --- & --- \\
            \rowcolor{lightgray}
            socket-benchmark & 1,209 & 5 & 4 & 
                       --- & --- &
                       --- & --- \\
            socket-throughput & 1,100 & 5 & 5 & 
                       --- & --- &
                       --- & --- \\
            % tcp2ext & 322 & 5 & 1 & 
            %            --- & --- &
            %            --- & --- \\
            \rowcolor{lightgray}
            tcp\_server & 750 & 2 & 2 & 
                       --- & --- &
                       --- & --- \\
            threaded-socket & 1,059 & 5 & 3 & 
                       --- & --- &
                       --- & --- \\
            % \midrule
            % Total & & & & 
            %           & &  
            %           & \\
		\bottomrule
	\end{tabular}
    % }
\end{table*}

\begin{tcolorbox} [colback=gray!20!white]
\textbf{Answer to RQ3:} 
\ccSysName is capable of identifying bad partitioning issues in real-world projects.
Through a review of 154 issues reported by \ccSysName in the wild, we have confirmed that 130 of them were true alarms by ourselves.
\end{tcolorbox}

\section{Threats to Validity} \label{s:val}

\subsection{Internal Validity}
We use the \ccBenchName dataset, which includes manually injected vulnerabilities, to evaluate \ccSysName.
This could lead to a discrepancy between the distribution of bad partitioning issues in the dataset and those in real-world TEE projects, thereby affecting the evaluation of detection accuracy.
It is possible that we incorporate more vulnerabilities from real-world TEE projects into \ccBenchName to reduce biases introduced by manual injection.

The performance of \ccSysName also heavily relies on the underlying static analysis tools.
However, most tools (\eg, CodeQL and Frama-C~\cite{framec}) typically necessitate successful compilation and expert customization.
This dependency limits their applicability to analyzing incomplete or uncompilable code.
If these tools fail to accurately generate data flows, the detection capabilities of \ccSysName could be compromised.
LLMDFA~\cite{wang2024llmdfa} proposed by Wang \etal makes up for this defect.
Combined with the large language model, LLMDFA effectively improves the accuracy of identifying data flow paths.

\subsection{External Validity}
The evaluations focus on TEEs based on ARM TrustZone. The detection performance of \ccSysName on other TEE architectures, such as Intel SGX, AMD SEV~\cite{10.1145/3623392, 10.1145/3214292.3214301} and RISC-V-based TEE~\cite{10.1145/3433210.3453112, 9343170}, needs further evaluation.
In addition, the evaluated projects are relatively small in scale and do not cover complex industrial-level projects.
Future studies should evaluate \ccSysName on a broader range of TEE architectures and large-scale real-world projects, thereby improving the external applicability of the experimental results.

\subsection{Construct Validity}
Our study defines three types of rules to identify bad partitioning issues.
However, in practical development, there may be certain security assumptions that violate these rules without actually compromising security, potentially leading to false positives.
For example, directly outputting a static string (\eg, a software version number, user id, etc.) that does not contain any sensitive data is considered secure and does not require encryption.
Therefore, if these definitions are not accurate or comprehensive enough, it may lead to biased detection results.
We need to update the rules based on the newly discovered bad partitioning issues in actual TEE applications to expand the detection capabilities of \ccSysName.

\section{Related Work} \label{s:related}
\subsection{Vulnerabilities in TEE}
TEE has become integral to security-critical applications by providing isolated execution environments to protect sensitive data and computations. However, TEE applications are still affected by some vulnerabilities, as demonstrated by extensive research across various TEE technologies.
Cerdeira \etal~\cite{9152801} conducted a study on vulnerabilities in TrustZone-based systems. They have revealed critical implementation bugs, such as buffer overflows and inadequate memory protection mechanisms. Their work also underscores the role of hardware-level issues like microarchitectural side-channels in compromising TEE security.
Fei \etal~\cite{10.1145/3456631} provided a taxonomy of secure risks in Intel SGX and discussed attack vectors, such as hardware side-channel exploits and software code vulnerabilities.
Similarly, Khandaker \etal~\cite{10.1145/3373376.3378486} introduced the concept of COIN attacks against SGX, and they expose the insecurity of untrusted interfaces in TEE projects.
Bulck \etal~\cite{10.1145/3319535.3363206} assessed the vulnerabilities at the level of the Application Binary Interface (ABI) and the Application Programming Interface (API) that can lead to memory and side-channel attacks in TEE runtime libraries.

To mitigate the risk of attacks on TEE projects, some research has designed tools to detect vulnerabilities within TEE.
Briongos \etal~\cite{10.1145/3627106.3627187} present CloneBuster for detecting cloning attacks on Intel SGX applications. CloneBuster can find whether there are same TEE application binaries running on one platform, which may roll back the TEE applications to the previous state. 
Ghaniyoun \etal~\cite{10.1145/3579371.3589070} designed TEESec, a pre-silicon framework for discovering microarchitectural vulnerabilities in TEE. TEESec can test an underlying microarchitecture against data and metadata leakage across isolation boundaries.

However, most of these studies focus on the security of TEE hardware architecture and development kits that may have side-channel attack vulnerabilities. Research addressing bad partitioning issues in TEE projects remains limited, despite it being an equally critical concern.
Poorly designed TEE code partitioning can result in severe security risks, such as the leakage of sensitive data or the injection of untrusted data, which could influence the integrity of TEE application execution.
Our work addresses this gap by comprehensively detecting security vulnerabilities caused by bad partitioning in TEE code, particularly in scenarios involving data interactions between the normal world and TEE with input, output, and shared memory. 

\subsection{TEE Application Partitioning}
Some research has attempted to mitigate vulnerabilities arising from bad partitioning by porting entire programs into TEE. 
For example, CryptSQLite~\cite{8946540} is a database system that places the entire database engine into TEE. 
This approach prevents attackers from stealing or tampering with critical data computed in TEE from the normal world, effectively protecting the confidentiality and integrity of data during database retrieval.
RT-TEE~\cite{9833604} is also a cyber-physical system that puts the whole flight control code into TrustZone. This can protect the stability of equipment like drones.
However, this way significantly increases the Trusted Computing Base (TCB) of the applications, contradicting the principle of maintaining a small TCB for trusted code~\cite{DBLP:conf/ndss/ShindeTTS17, 10.1145/3313808.3313810}.

To address this, some other research made efforts to identify sensitive data and functions that require protection and place them inside TEE, ensuring a balance between security and efficiency.
Rubinov \etal~\cite{10.1145/2884781.2884817} developed an automatic approach for partitioning Android applications into critical code running in TEE and client code in the normal world.
They also reduce the overhead due to transitions between the two worlds by optimizing the granularity of TEE code.
Sun \etal~\cite{10577323} proposed dTEE that enables developers to declare tiered-sensitive variables and functions of existing IoT applications. What's more, dTEE can automatically transform device drivers into trusted ones.
Moreover, DarkneTZ~\cite{10.1145/3386901.3388946} is a framework that limits the attack surface of the model against Deep Neural Networks with the help of Trusted Execution Environment in an edge device.
They partitioned the model into sensitive layers inside TEE, and other layers executed in the normal world. 
This will provide reliable model privacy and defend against membership inference attacks.

The above TEE program partitioning schemes ensure software security while also maintaining the efficiency of program execution.
However, as discussed in Section~\ref{s:bp}, improper TEE partitioning can still result in vulnerabilities. Therefore, a tool capable of analyzing and identifying bad partitioning issues in TEE implementations is essential.

\section{Conclusion} \label{s:con}
Bad partitioning in TEE projects can lead to vulnerabilities during data transmission between the normal world and TEE. 
To address this problem, we designed \ccSysName, a tool that can identify security risks arising from bad partitioning by analyzing three types of data interactions: output, input, and shared memory. 
Additionally, we created the first partitioning errors benchmark, \ccBenchName, consisting of 110 test cases to evaluate the effectiveness of \ccSysName.
From the evaluation results, \ccSysName can help to find bad partitioning issues in TEE applications.
In the future, we aim to explore automated solutions for fixing bad partitioning issues in TEE applications based on the detection results.

% or


% \appendices

% \section{}

\section*{Acknowledgment}
This research is supported by the National Research Foundation, Singapore, and the Cyber Security Agency of Singapore under its National Cybersecurity R\&D Programme (Proposal ID: NCR25-DeSCEmT-SMU). Any opinions, findings and conclusions or recommendations expressed in this material are those of the author(s) and do not reflect the views of the National Research Foundation, Singapore, and the Cyber Security Agency of Singapore.


\ifCLASSOPTIONcaptionsoff
  \newpage
\fi



%\bibliographystyle{IEEEtran}


% \end{thebibliography}

\balance
\bibliographystyle{IEEEtran}
\bibliography{bibtex/bib/sample}


% \end{IEEEbiography}


%\newpage



%\vfill


\end{document}